\pdfoutput=1

\documentclass[12pt]{article}

\usepackage[letterpaper,margin=1.25in]{geometry}
\usepackage{graphicx}
\usepackage{amsmath}
\usepackage[hidelinks]{hyperref}
\usepackage{enumerate}
\usepackage{subcaption}
\usepackage{authblk}

\title{\texttt{blackbox:} A procedure for parallel optimization of expensive black-box functions}

\author[]{Paul Knysh\thanks{paul.knysh@gmail.com} }
\author[]{Yannis Korkolis\thanks{yannis.korkolis@unh.edu, Corresponding author}}

\affil[]{University of New Hampshire, USA}

\date{}

\begin{document}

\maketitle

\begin{abstract}

\noindent This note provides a description of a procedure that is designed to efficiently optimize expensive black-box functions. It uses the response surface methodology by incorporating radial basis functions as the response model. A simple method based on a Latin hypercube is used for initial sampling. A modified version of CORS algorithm with space rescaling is used for the subsequent sampling. The procedure is able to scale on multicore processors by performing multiple function evaluations in parallel. The source code of the procedure is written in Python.


\end{abstract}

\section{Introduction}

Engineering problems often involve search for optimal parameters of physical objects (geometry, chemical composition etc) or mathematical models (coefficients, input data sets etc). Considering the possible complexity of these problems, it is not always easy or even possible to come up with an analytical approach that answers the question. Instead, trial-and-error search using computer simulation can be performed, which, if done manually, is tedious and inefficient, if feasible at all.

One can build a code interface (function/procedure) that has input --- set of trial values, and output --- some scalar value that represents cost or error. This value is calculated automatically every time the simulation is performed. From this point of view, the problem is reduced to one of mathematical optimization. However, the objective function does not have an analytical form and, what is more crucial, is usually expensive (it can take many hours to evaluate for a single set of input parameters), and therefore is called expensive black-box function.

The procedure proposed here allows to perform efficient optimization of expensive black-box functions. Usage of the response surface methodology \cite{box1951experimental} based on radial basis functions \cite{powell1990theory} allows us to reconstruct (and subsequently optimize) a given black-box function with a limited number of function evaluations. The initial stage of the procedure is based on a custom Latin hypercube sampling \cite{mckay2000comparison}. Subsequent stage is based on a modification of CORS algorithm \cite{regis2005constrained} with space rescaling. In addition, the procedure is designed to scale on multicore processors by mapping the function on sets of arguments in a parallel way (each core is running its own function evaluation), which results in a speedup that is equal to a number of cores available.

\section{Overview of the procedure}

The goal of the procedure is to efficiently minimize a non-negative function $f$, with each of its variables having a specified independent range of values. To maximize $f$, the procedure can be simply applied on $\frac{1}{f+1}$ or similar expression.

The procedure proposed here consists of four main steps, which are described in detail in the following sections:

\begin{enumerate}
\item
Rescaling of variables
\item 
Initial sampling
\item
Rescaling of the objective function
\item
Subsequent iterations
\end{enumerate}

\subsection{Rescaling of variables}

Since further analysis will be based on distances between sampled points, it is vital to ensure that the ranges of variables are not very different. For example, the variables could have different physical nature and might exhibit differences by orders of magnitude. Therefore, the natural step is to normalize each range into a range of $[0,1]$. For a variable $v_i$ that is in range $[a_i,b_i]$ the following simple transformation is used:
\begin{equation}
\xi_i = \frac{v_i - a_i}{b_i - a_i} \text{.}
\end{equation}
After such rescaling, the search space becomes simply a unit cube\footnote{This should be understood as a multidimensional entity. For convenience, the prefix 'hyper' is not used, but is assumed by default for any geometry --- cube, ball, plane etc}.

\subsection{Initial sampling}

\begin{figure}[!htb]
\centering
\begin{subfigure}{0.3\columnwidth}
  \centering
  \includegraphics[width=\columnwidth]{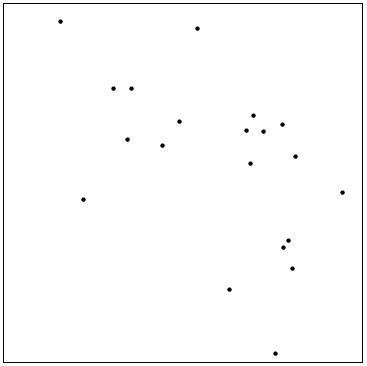}
  \caption{}
  \label{fig:lh_random}
\end{subfigure}
\hfill
\begin{subfigure}{0.3\columnwidth}
  \centering
  \includegraphics[width=\columnwidth]{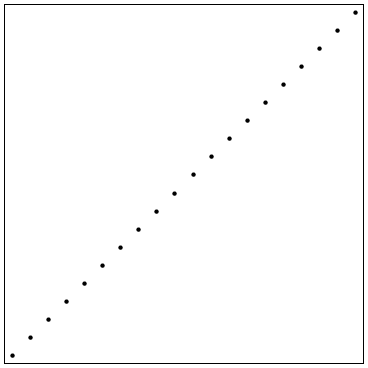}
  \caption{}
  \label{fig:lh_initial}
\end{subfigure}
\hfill
\begin{subfigure}{0.3\columnwidth}
  \centering
  \includegraphics[width=\columnwidth]{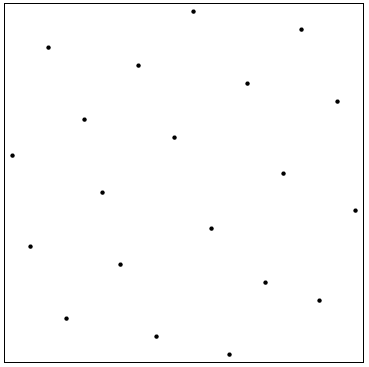}
  \caption{}
  \label{fig:lh_final}
\end{subfigure}
\caption{Placing $20$ points into a unit square. (a) --- typical placement of random points, (b) --- initial LH, (c) --- eventual LH.}
\end{figure}

The quality of the response surface that reconstructs a black-box function depends significantly on the initial set of samples. A uniform mesh, while acceptable for lower dimensions, becomes inefficient for higher dimensions since the number of evaluations grows exponentially with the space size. Generating random samples allows the number of samples to be independent of the space size, but the samples will not cover the search space uniformly (2D example is in Fig.~\ref{fig:lh_random}).

The problem of placing a given amount of points $n$ into a unit cube in a uniform manner can be solved with the Latin hypercube (LH) \cite{mckay2000comparison}. Essentially, the $n$ points are placed at the nodes of a uniform mesh of the same size in such a way that there is exactly one point in each axes-aligned plane containing it.

A simple method for constructing a LH of a reasonable uniformity is presented here. At first all samples are placed diagonally (2D example is in Fig.~\ref{fig:lh_initial}), which forms the initial LH. Subsequently, two random, axes-aligned planes are picked and exchanged. If the LH obtained has an improved uniformity of samples, it is kept, otherwise another random exchange is performed until an improvement is achieved. The measure of uniformity, or spread (which has to be minimized), used here can be introduced in the following way:
\begin{equation}
S =  \sum_i^n \sum_{j(>i)}^n \frac{1}{\| x_i - x_j \|}\text{,}
\end{equation}
where $\| x_i - x_j \|$ is the distance between two points. By repeating such exchanges iteratively a reasonable number of times (for example, $1000$), a LH with a uniform placement of samples can be obtained (2D example is in Fig.~\ref{fig:lh_final}).

\subsection{Rescaling of the objective function}

The raw values of the objective function might be misleading and hard to interpret (for example, value of $1.234\times 10^5$ might be high or low depending on reference to compare with). To overcome this problem, function values are rescaled into the $[0,1]$ range. This makes it much easier to interpret the values relative to $1$ (the worst value) and $0$ (the best value).

Also, sometimes outliers in the form of extremely high values of the objective function may appear. These values, if kept, can pollute the overall shape of the response surface. To eliminate such possibility, a specified fraction of samples with lowest values obtained on the initial stage is kept (and corresponding threshold value $t$ is introduced), while the rest of the values (that are higher than $t$) are discarded.

The rescaled function can be defined in the following way:
\begin{equation}
f^* = 
\begin{cases}
\frac{f}{t}, \, f<t \\
1, \, f \geq t
\end{cases} \text{.}
\end{equation}

\subsection{Subsequent iterations}
\subsubsection{Modified CORS algorithm}

After the initial sampling, a response surface using cubic radial basis functions (RBF) can be constructed. The following expression was used \cite{holmstrom2008adaptive}:
\begin{equation}
\label{eqn:rbf}
s_n(x)= \sum_{i=1}^n \lambda_i \phi(\| x-x_i \|) + b^T x + a
\text{,}
\end{equation}
where $s_n$ is the response surface reconstructed with $n$ sampled points ($x_i$), $\phi$ is a cubic function ($\phi(r)=r^3$) and $\lambda_i$, $b$, $a$ are coefficients that are determined from the fact that the response surface interpolates all samples $x_i$. The RBF fit, once constructed, is able to predict the value of the objective function at an arbitrary point $x$.

Subsequent iterations are performed in accordance with the modified CORS algorithm \cite{regis2005constrained}. The concept of this algorithm is the following --- while the current fit is minimized, a new sampled point has to be not closer than $r$ to each of the previously sampled points. The value of $r$ loops over a set of values in a periodic way, which prevents the algorithm from being trapped in a local minimum.

The procedure described here uses the following interpretation of CORS algorithm --- a ball of radius $r$ is placed around each of previously sampled points and the new point is required to minimize the current fit, but to be outside of every ball. Such analogy allows us to introduce a density of these balls as the total volume of the balls divided by the total volume of a unit cube. This density can be assumed to start with an initial value and decrease (decay) with iterations with a given rate. By controlling this initial density and rate of decay, different search strategies (for example, with more attention to global/local search) can be achieved.

Since the volume of a unit cube is $1$, the following can be written:
\begin{equation}
\rho \leq Nv \text{,}
\end{equation}
where $\rho$ is the density of the balls, $N$ is amount of previously sampled points, $v$ is a volume a single ball. The inequality indicates that the actual density can be less than $Nv$ --- some balls may intersect with each other or may have some fraction of their volume outside of the cube.

The volume of a $d$-dimensional ball is equal to $v_1 r^d$, where $v_1$ is a volume of a ball with radius $1$. The current amount of balls $N$ is equal to $n+i-1$, where $n$ is the amount of initial samples and $i$ is the number of current subsequent iteration (counted from $1$). Then:
\begin{equation}
\rho \leq (n+i-1) v_1 r^d \text{,}
\end{equation}
\begin{equation}
\label{eqn:r_rho}
r \geq \left( \frac{\rho}{(n+i-1)v_1} \right)^{1/d} \text{.}
\end{equation}
The volume $v_1$ can be expressed with the gamma function:
\begin{equation}
\label{eqn:v_1}
v_1 = \frac{\pi^{\frac{d}{2}}}{\Gamma \left( \frac{d}{2} + 1 \right)} \text{.}
\end{equation}
The density can then be introduced in the following form:
\begin{equation}
\label{eqn:rho}
\rho = \rho_0 \left( \frac{m-i}{m-1} \right)^p \text{,}
\end{equation}
where $\rho_0$ is initial density, $p$ is a rate of density decay, $m$ is a total number of subsequent iterations. If $p=1$, then density linearly decays with iterations, if $0<p<1$ --- slower than linearly, if $p>1$ --- faster than linearly.

Equation \eqref{eqn:r_rho} (with equality sign) together with \eqref{eqn:v_1} and \eqref{eqn:rho} allows us to find the current radius $r$ on every subsequent iteration.

\begin{figure}[!htb]
\centering
\begin{subfigure}{0.3\columnwidth}
  \centering
  \includegraphics[width=\columnwidth]{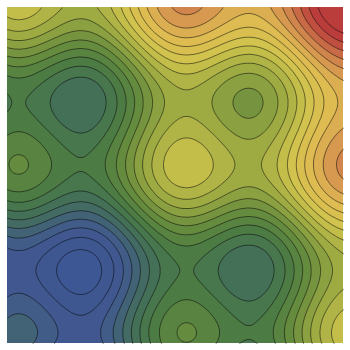}
  \caption{}
  \label{fig:test_f}
\end{subfigure}
\hfill
\begin{subfigure}{0.3\columnwidth}
  \centering
  \includegraphics[width=\columnwidth]{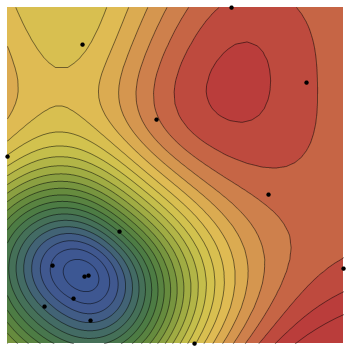}
  \caption{}
  \label{fig:test_rough}
\end{subfigure}
\hfill
\begin{subfigure}{0.3\columnwidth}
  \centering
  \includegraphics[width=\columnwidth]{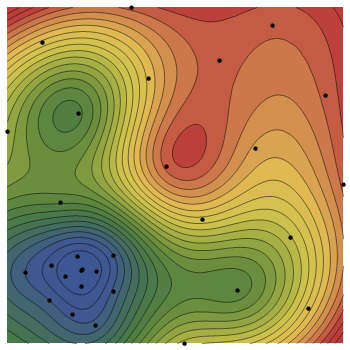}
  \caption{}
  \label{fig:test_fine}
\end{subfigure}
\caption{(a) --- function given on $0 \leq x,y \leq 1$, (b)/(c) --- results of running a procedure using 15/30 function evaluations.}
\label{fig:test}
\end{figure}

An application example of the proposed procedure is shown in Fig.~\ref{fig:test}. The function $f(x,y)= \cos 4 \pi x + \cos 4 \pi y + 5(x+y)+2, \, 0 \leq x,y \leq 1$ shown in Fig.~\ref{fig:test_f} is assumed to be unknown and can only be evaluated at given points. Notice that the function has 1 global and 3 local minima. Fig.~\ref{fig:test_rough} shows the result of applying the procedure using 15 function evaluations (10 on initial stage and 5 on subsequent) and Fig.~\ref{fig:test_fine} --- using 30 evaluations (20 on initial stage and 10 on subsequent). Function evaluations are represented with black dots. It can be seen that the quality of the response surface depends significantly on the number of function evaluations selected. While in either case it was possible to identify the location of the global minimum, some of the local minima were not captured.

\subsubsection{Space rescaling}

Oftentimes the global minimum is located at the bottom of a valley-like feature. In this case, the original function may be represented inaccurately by its RBF fit, which will result in poor accuracy of the resulting optimum that the procedure will yield. A way to improve the quality of the RBF fit based on space rescaling is discussed here.

First, the RBF expression before rescaling \eqref{eqn:rbf} is used to construct the initial fit. After that, a large number of random samples (typically $10000$ or more) is populated in the unit cube and some fraction of best points (typically $5\%$ of total number) is selected based on their RBF values. This 'cloud' of points roughly approximates the shape of the valley feature and is used for the space rescaling procedure.

\begin{figure}[!htb]
\centering
\begin{subfigure}{0.3\columnwidth}
  \centering
  \includegraphics[width=\columnwidth]{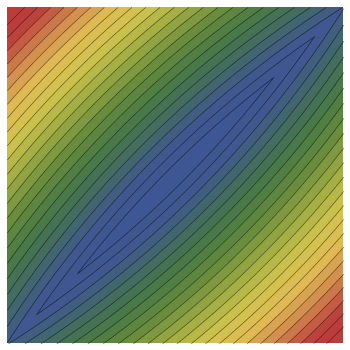}
  \caption{}
  \label{fig:rbf_f}
\end{subfigure}
\hfill
\begin{subfigure}{0.3\columnwidth}
  \centering
  \includegraphics[width=\columnwidth]{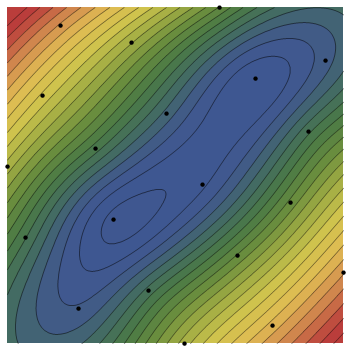}
  \caption{}
  \label{fig:rbf_unscaled}
\end{subfigure}
\hfill
\begin{subfigure}{0.3\columnwidth}
  \centering
  \includegraphics[width=\columnwidth]{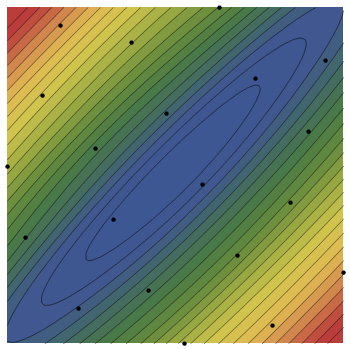}
  \caption{}
  \label{fig:rbf_scaled}
\end{subfigure}
\caption{(a) --- function given on $0 \leq x,y \leq 1$, (b) --- original RBF fit \eqref{eqn:rbf}, (c) --- RBF fit with space rescaling \eqref{eqn:rbf_T}.}
\end{figure}

The covariance matrix of the cloud can be evaluated as:
\begin{equation}
C=\text{cov} (\text{cloud})  \text{.}
\end{equation}
Then eigensystem of the covariance matrix is found as:
\begin{equation}
\alpha_i, m_i = \text{eig}(C) \text{,}
\end{equation}
where $\alpha_i$ are the eigenvalues and $m_i$ are the eigenvectors. Then, a scaling matrix is introduced as:
\begin{equation}
T=\begin{pmatrix}
m_1 / \sqrt{\alpha_1} \\
m_2 / \sqrt{\alpha_2} \\
\vdots \\
m_d / \sqrt{\alpha_d}
\end{pmatrix}
\text{}
\end{equation}
and the RBF expression \eqref{eqn:rbf} is updated:
\begin{equation}
\label{eqn:rbf_T}
s_n(x)= \sum_{i=1}^n \lambda_i \phi(\| T(x-x_i) \|) + b^T x + a
\text{.}
\end{equation}
To provide an example of the procedure discussed, the simple function $f(x,y)=|x-y|+\left( \frac{x+y-1}{3} \right)^2, 0 \leq x,y \leq 1$ shown in Fig.~\ref{fig:rbf_f} is assumed to be unknown and can only be evaluated at given points. The original RBF fit reconstructed with 20 samples (black dots) is shown in Fig.~\ref{fig:rbf_unscaled}. An RBF fit with space rescaling is shown in Fig.~\ref{fig:rbf_scaled}. It can be seen that the rescaling procedure improves the quality of the fit significantly.

\section{Using multiple cores}

The proposed procedure is designed to scale on multicore processors to handle expensive black-box functions more efficiently. This is achieved simply by dividing the samples on the initial and subsequent stages into batches that are evaluated simultaneously. The size of the batch is equal to the number of cores available. The Python package \texttt{multiprocessing} and its method \texttt{map} were used in the code to map a given black-box function on a batch of samples in a parallel way. The resulting speedup is equal to the number of cores available.

\section{Conclusions}

A procedure that is able to efficiently optimize expensive black-box functions is described. It is based on the response surface methodology, uses a powerful sampling technique and is able to scale on multicore processors, all of which allow us to locate the global optimum with a limited number of function evaluations.

\section{Acknowledgements}

This work was performed under the U.S. National Science Foundation CAREER award (grant CMMI-1150523). This support is acknowledged with thanks.

\end{document}